\documentclass[aps,12pt,floatfix,final,notitlepage,oneside,onecolumn,nobibnotes,
nofootinbib,superscriptaddress,noshowpacs,centertags]{revtex4}

\RequirePackage{amssymb,amsmath} \RequirePackage{eufrak} \RequirePackage{mathtext}
\RequirePackage[cp1251]{inputenc} \RequirePackage[T2A]{fontenc}
\RequirePackage[russian,english]{babel} \RequirePackage{bm} \RequirePackage{array}
\RequirePackage{longtable} \RequirePackage{graphicx} \RequirePackage{calc}
\RequirePackage{ifthen} \RequirePackage{caption2} \RequirePackage{subfigure}

\def\refitem#1{\relax}

\begin{document}

\selectlanguage{english}

\title{Gas density and star formation in the rarified regions of discs of normal and LSB galaxies}

\author{\firstname{~A.~V.} \surname{Zasov}} \author{\firstname{~O.~V.} \surname{Abramova}}
\affiliation{Sternberg Astronomical Institute of the Moscow State University, Moscow,
Russia}

\begin{abstract}
We calculated the radial profiles of the azimuthally averaged midplane gas volume density
$\rho_g$ for 11 high surface brightness (HSB) spiral galaxies, 7 low surface brightness
(LSB) galaxies and 3 S0 galaxies assuming their gaseous layers to be in the equilibrium
state in the plane of marginally stable stellar discs. We compared the surface star
formation rate ($\Sigma_{SFR}$) and star formation efficiency
($SFE=\Sigma_{SFR}/\Sigma_{gas}$) with $\rho_g$ and stellar surface density $\Sigma_s$
assuming the latter to be proportional to disc surface brightness. Both HSB and LSB
galaxies follow a single sequence $\Sigma_{SFR}-\rho_{g}$ and $SFE-\Sigma_s$ or
$SFE-\rho_s$. It means that the conditions of star formation are similar in the outer
discs of normal spiral galaxies and in the inner regions of LSB galaxies if their
stellar discs have similar densities. The relationship between $SFE$ and $\rho_s$ is close
to the law $SFE$ $\sim \rho_s^{1/2}$ expected in the theoretical model of self-regulated
star formation proposed by Ostriker et al.\cite{OstrikerAll10}. The alternative
explanation is to propose that $SFE$ is proportional to a frequency of vertical
oscillation of gas clouds around the disc midplane. In the most rarified regions of LSB
galaxies the efficiency of star formation is nearly independent on gas and stellar disc
densities being higher in the mean than it is expected from the extrapolation of the power
law fit for HSB sample galaxies. Evidently in these regions with extremely low $\rho_{g}$
SFE depends on local density fluctuations rather than on the azimuthally averaged disc
parameters.
\end{abstract}

\maketitle

\section{Introduction}

\hspace{0.6cm}Star formation is the main process governing the evolution of galaxies. Star
formation rate ($SFR$) is connected with the amount of gas in  galactic discs, although
the relationship between $SFR$ and gas mass or gas density  is   ambiguous  and poorly
understood. It is clear that local gas density, averaged over large enough area, or
azimuthally averaged surface density at a given radial distance $R$ is not the only factor
which determines the current $SFR$. Indeed, $SFR$ may be different  in the inner regions
and in the outer regions of galactic disks even for similar surface density of gas because
local properties of stellar disc and interstellar medium vary along the radius: a
thickness of gas layer increases, a density and  internal pressure of gas decreases
parallel with its metallicity. Some dynamical parameters such as the angular velocity of
rotation also change along the radius.  Despite all these complexities, there exist
simple, although not very tight, empirical relationships, known as the Schmidt or
Schmidt-Kennicutt laws or their different modifications, linking the rate of star
formation per unit surface area of a disc $\Sigma_{SFR}$ (both local or azimuthally
averaged one) with a surface density of gas $\Sigma_g$ at a given distance $R$ from the
center (see f.e.~\cite{Kennicutt98, Wong&Blitz02, BoissierAll03, LeroyAll08, BigielAll08,
SchrubaAll11} and references therein). In a wide range of $\Sigma_g$ the relationship may
be written as $\Sigma_{SFR}\sim\Sigma_g^N$, where $N\approx 1.5$. The value of $N$ has a
tendency to be higher (the slope becomes steeper) for the atomic gas-dominated outer disc
regions, although the scatter of points is large there.

A special problem is to explain how do stars form in the conditions of very low surface
density of gas which exists at the peripheral regions of disks of high surface brightness
(HSB) spiral galaxies, in some gas-poor lenticular galaxies and in low surface brightness
(LSB) galaxies. In all these cases the surface density of gas is too low for the
development of gravitational and/or  thermal gas instabilities (at least for the usually
adopted properties of interstellar medium). Nevertheless, as UV observations of GALEX  and
H$\alpha$ imaging showed, a presence of young stellar population in many spiral galaxies
is noticeable up to the optical radius $R_{25}$ and even beyond (see~\cite{Bigiel_etal10,
GoddardAll10} and references therein). The origin of the fireplaces of star formation
there remains puzzling.

To describe how favorable are the existing conditions for the current formation of stars
it is convenient to use the efficiency of star formation $SFE$, that is a star formation
rate per unit of gas mass: $SFE=\Sigma_{SFR}/\Sigma_g$, or gas consumption time
$\tau=SFE^{-1}$. As observations show, $SFE$ monotonically decreases
along the radius parallel with the gas or stellar surface densities -- at least in the
outer parts of galaxies~\cite{Wong&Blitz02, BigielAll08, Bigiel_etal10, LeroyAll08, KoopmannAll06,
ShiAll11}. Note that the relationship between $\Sigma_{SFR}$ and $\Sigma_g$ or
$\Sigma_\textrm{HI}$ in the outer parts of discs strongly differs for different galaxies.
In some cases there is a break in radial profile of UV or H$\alpha$ brightness, expected
in theoretical models (see f.e. Goddard et al.~\cite{GoddardAll10}), but in some galaxies
the slope remains nearly constant down to extremely low gas density. A crucial role in
star formation in low density regions may belong to a heating of gas by newly formed stars
or some outer sources: the indirect evidence of their presence is that the line-of-sight
velocity dispersion of HI remains high enough (5-10 km/s) even at  the far peripheries of
galaxies~\cite{TamburroAll09}.

Observations of CO emissions give evidence that the current $SFR$ is in general linearly
proportional to the molecular gas surface density $\Sigma_\textrm{mol}$ down to a several
solar masses per pc$^2$, although the scatter remains high~\cite{BigielAll11,
SchrubaAll11}. It means that $SFE$, as we define it, just reflects the fraction of
molecular gas, or, in other words, a condition of formation and survival of molecular
clouds -- predecessors of young stars. A correlation of $\Sigma_{SFR}$ with neutral gas
density $\Sigma_\textrm{HI}$ is not so well defined as that with molecular gas, but it
really exist, especially  in the outer regions of discs, where both $\Sigma_\textrm{mol}$
and $\Sigma_\textrm{mol}/\Sigma_\textrm{HI}$ are low. For LSB galaxies, a comparison of
total $SFR$ with the total mass of gas shows that $SFR$ is significantly below the
extrapolated power law fit applied to the HSB sample (see~Wyder et al\cite{WyderAll09}).

The quantitative analysis and interpretation of these relationships is complicated by the
fact that the surface gas density is the integral of volume density along the line of
sight, and, hence,  even after correction for disc projection, it depends on the thickness
of gas layer which changes significantly along the radius of a given galaxy. This
circumstance is often ignored, which may lead to significant systematic errors. As we
showed in the previous paper (Abramova, Zasov~\cite{AbrZas11}), if to replace the surface
gas density $\Sigma_g$ by the midplane volume gas density $\rho_g$ calculated in the frame
of axisymmetric equilibrium disc model, the relationship between $\Sigma_{SFR}$ and
$\rho_g$ becomes more tight and well defined (see Figures 1a,b in this paper) which proves
that fundamental relationship is between the $\Sigma_{SFR}$ and gas volume density, not
the surface one.

In this paper we try to compare the efficiency of star formation $SFE$ and its dependence
on the volume midplane density of gas and stars in spiral galaxies, LSB galaxies and a few
S0-galaxies with noticeable star formation.

\section{Midplane gas densities}

\hspace{0.6cm}The method of estimation of midplane gas and star densities $\rho_g$,
$\rho_s$  was described earlier (see Abramova, Zasov~\cite{AbrZas11, AbrZas08}). We used
the velocity curves of galaxies  and the radial distributions of  brightness and surface
gas densities taken from the literature as the input data. The inner regions of galaxies
($R<1-3$~kpc) were not considered – mainly to avoid the influence of a bulge or a bar. The
main assumptions we accepted for calculations are counted below:
\begin{itemize}
\item gaseous layer and exponential stellar disc are assumed to be axisymmetric;
\item gaseous layer and stellar disc are in the vertical equilibrium state situated in the gravitational field of all components of a galaxy (stellar disc, HI and H$_2$ layers, and spherical pseudo isothermal dark halo);
\item gas velocity dispersion is taken to be $\sim$9~km/s for HI (if the direct estimations are absent) and 6~km/s for H$_2$ in HSB and S0 galaxies;
\item the local values of vertical stellar velocity dispersion assumed to be proportional to
the critical radial velocity dispersion for marginally stable discs which stems from the
numerical experiments (see Zasov et al.~\cite{ZKT03}).
\end{itemize}

The last assumption of marginal stability of stellar disc is usually valid for spiral
galaxies -- at least within several radial scalelengths, (see the arguments
in~\cite{ZasKhopSab10} and references therein), although its validity for the outer discs
and for LSB galaxies is questionable (see the discussion in~\cite{AbrZas11}). At any case,
if the disc of a galaxy is actually far from marginal stability (that is, a disc is
overheated and hence is thicker than expected), then the meanings of $\rho_g$ and $\rho_s$
obtained by the method we used may be considered as the upper limits.  It is worth to note
that the estimations of these densities are not sensitive to the adopted values of local
surface density of disc $\Sigma_d$ (which is usually close to $\Sigma_s$). A reason of it
is very simple: for marginally stable self-gravitating discs the increasing of $\Sigma_d$
leads to the proportional increasing of velocity dispersion of stars (or gas, if its
surface density is higher). In turn, it leads to the increasing of disc thickness $h$
proportional to $\Sigma_d$, and, as a result, the midplane volume density, being equal to
to $\Sigma_d/2h$, remains nearly the same.

\begin{figure}
\centering \mbox{\subfigure[]{\includegraphics[scale=0.40]{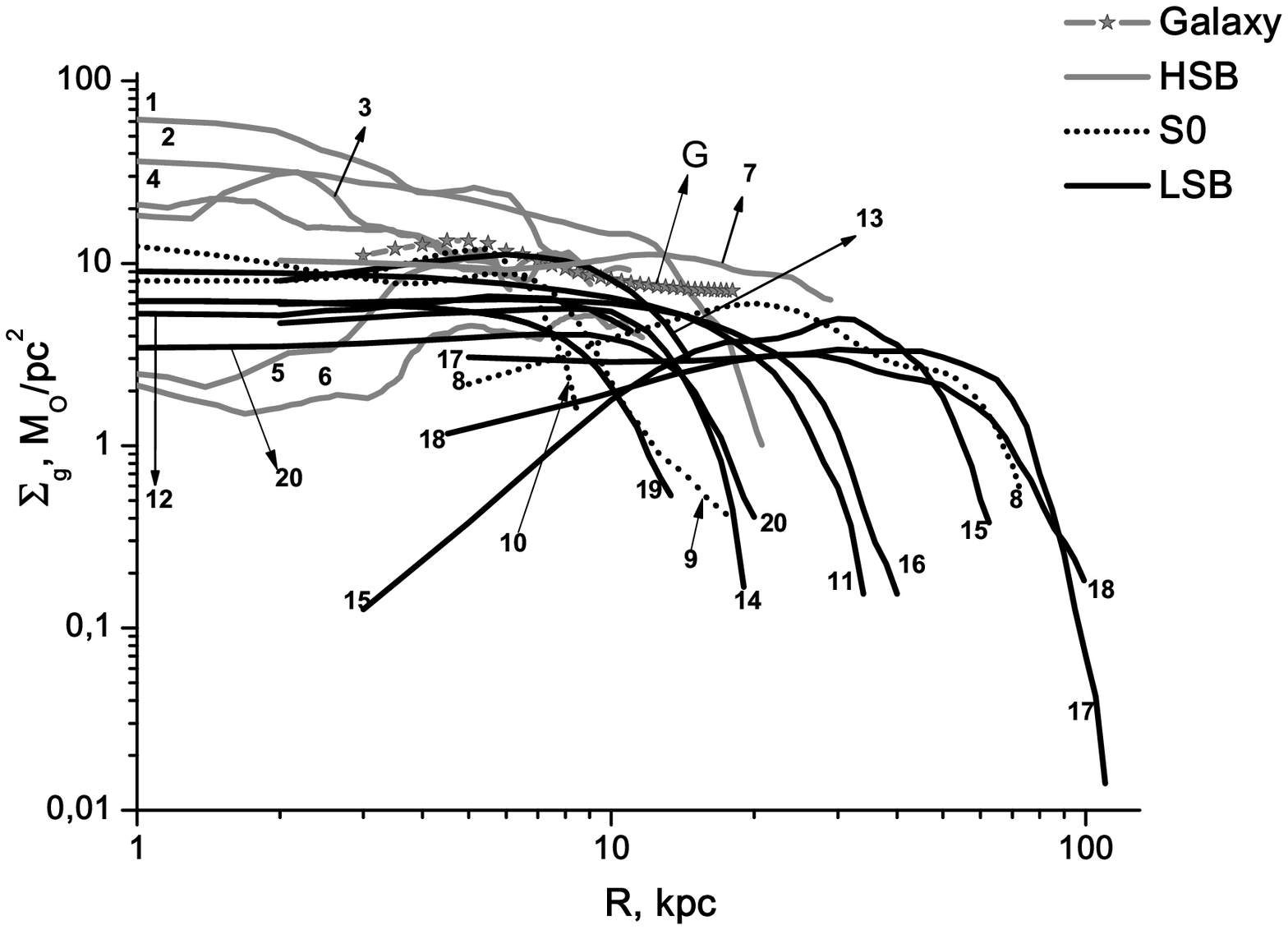}}\quad
\subfigure[]{\includegraphics[scale=0.40]{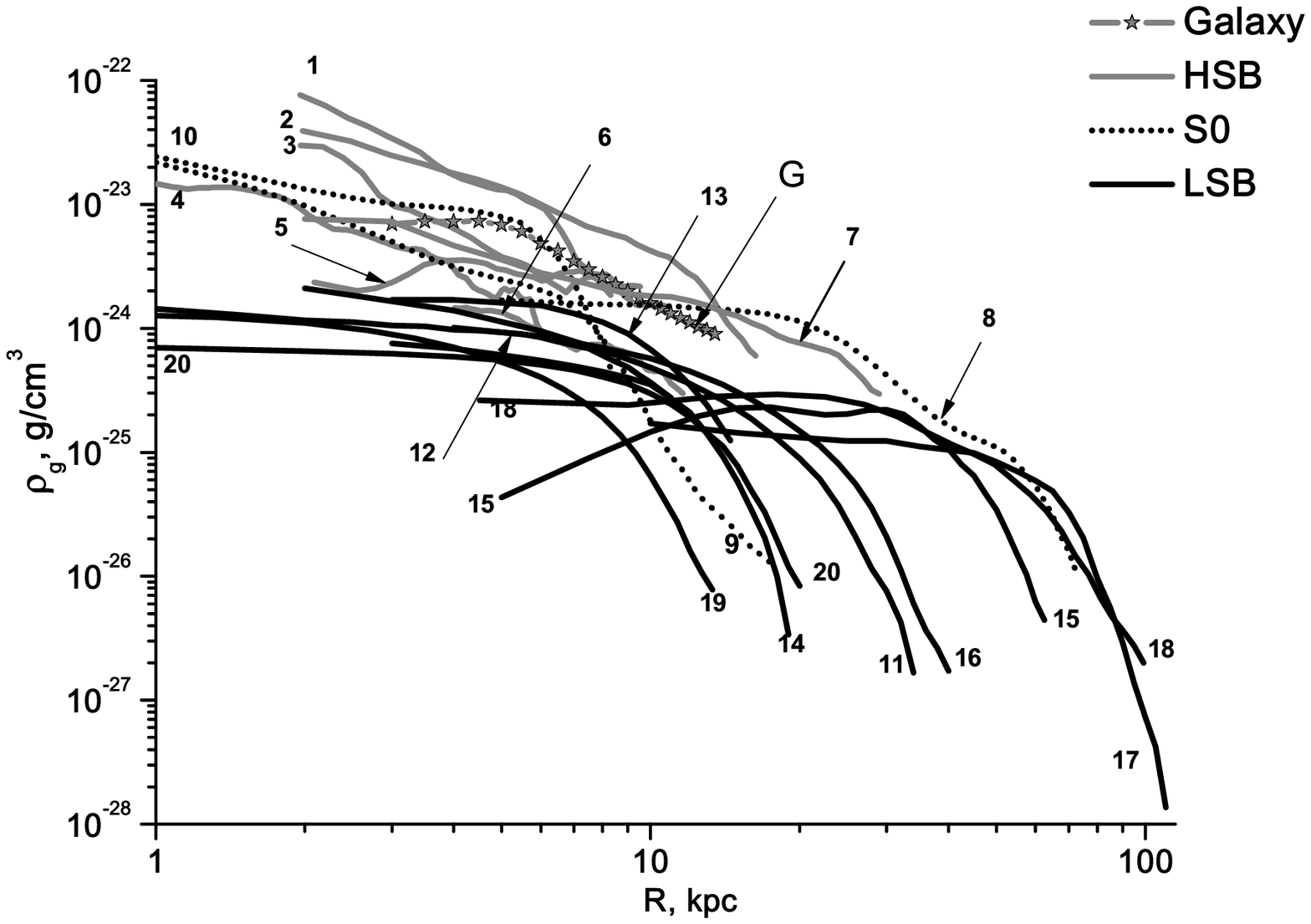}}} \caption{\footnotesize
Radial distribution of surface (a) and volume (b) gas densities for 21 galaxies of
different types. {\it Designations}. HSB galaxies: G - Galaxy, 1 - M51, 2 - M100, 3 -
M101, 4 - M33, 5 - M106, 6 - M81, 7 - ngc 289; S0: 8 - ugc 2487, 9 - ugc 11670, 10 - ugc
11914; LSB galaxies: 11 - ugc 1230, 12 - F568-3, 13 - F568-1, 14 - F568-v1, 15 - ugc 6614,
16 - ugc 128, 17 - Malin 1, 18 - Malin 2, 19 - F561-1, 20 - F574-1.}\label{fig1}
\end{figure}

Figs~\ref{fig1}(a,b) illustrate the radial  profiles of the azimuthally averaged surface
density $\Sigma_{SFR}$ (a) and  the volume density
$\rho_g(R)$ (b) calculated for 21 galaxies (7 HSB, one HSB + extended LSB disk (NGC~289), 10 LSB and 3 S0). The names of all galaxies  are counted
in the legend to Fig.~\ref{fig1}. References to the sources of the disc parameters we used
and the rotation curves may be found in Abramova, Zasov~\cite{AbrZas11, ZasAbr08}. We
assumed $H_0=75$~km/s/Mpc.
%%++ Figure:1

As one can expect, $\rho_g(R)$ decreases steeper than $\Sigma_g(R)$ due to the flaring of
gas layer. A mean volume density of the observed HI in the outskirt of LSB galaxies is
exremely low -- down to 10$^{-27}$~g/cm$^3$, that is three orders of magnitudes lower than
in the solar circle!

\section{Star formation rate and star formation efficiency}

\hspace{0.6cm}To match the midplane gas density with the observed star formation we
considered 7 normal galaxies (M33, M51, M81, M100, M101, M106 and Galaxy) and 8 LSB
galaxies (F561-1, F574-1, F568-1, F568-v1, F568-3, F568-6 (Malin~2), Main~1 and ugc~6614).
The references to the sources of data may be found in~\cite{AbrZas11}. Radial profiles of
$SFR$ based on the UV and/or far IR observations for 6 HSBs (except the Galaxy) were
calculated as  described in~\cite{ZasAbr06}, the necessary photometric data were taken
from~Boissier et al~\cite{BoissierAll04}. Radial profile of $\Sigma_{SFR}$ for our Galaxy
was taken from the far IR profile $L_{FIR}(R)$~\cite{BronfmanAll00} which was normalized
to $SFR=4\cdot10^{-9}$~M$_\odot$/(yr$\,$pc$^2)$ in the Solar neighborhood (it
corresponds to total $SFR_{tot}=3,6$~M$_\odot$/yr).

$FUV$ and $NUV$ profiles based on the GALEX observations which were used to calculate
radial profiles of $\Sigma_{SFR}$ for 8 LSB galaxies were presented by Wyder et
al.~(2009)~\cite{WyderAll09}. The most reliable and widely used method of estimation of
star formation rate is based on the combination of UV (or emission lines) intensity and
far infrared brightness to take into account the light absorption. However in the outer
regions of galaxies and in LSB galaxies the absorption is low, so the pure UV brightness
may  be taken as an indicator of $SFR$~\cite{WyderAll09}; in general this method gives the
bottom estimation of $SFR$. For LSB galaxies the molecular gas was ignored due to its low
content.

It should be remembered that the absolute estimates of $SFR$ are always very approximate
and may contain systematic errors, mostly due to the  uncertainties of the adopted stellar
IMF and the difficulties of taking into account the dust extinction. All indicators of
star formation relate mainly to stars of high and intermediate masses. Happily, there  are
no direct evidences that IMF strongly differs in the inner and in the outer regions of
galactic discs~\cite{YasuiAll08, GoddardAll10} or in LSB galaxies~\cite{WyderAll09}, which
allows to use the existing estimates of $SFR$ for comparison purposes.

\begin{figure}
\centering \mbox{\subfigure[]{\includegraphics[scale=0.30]{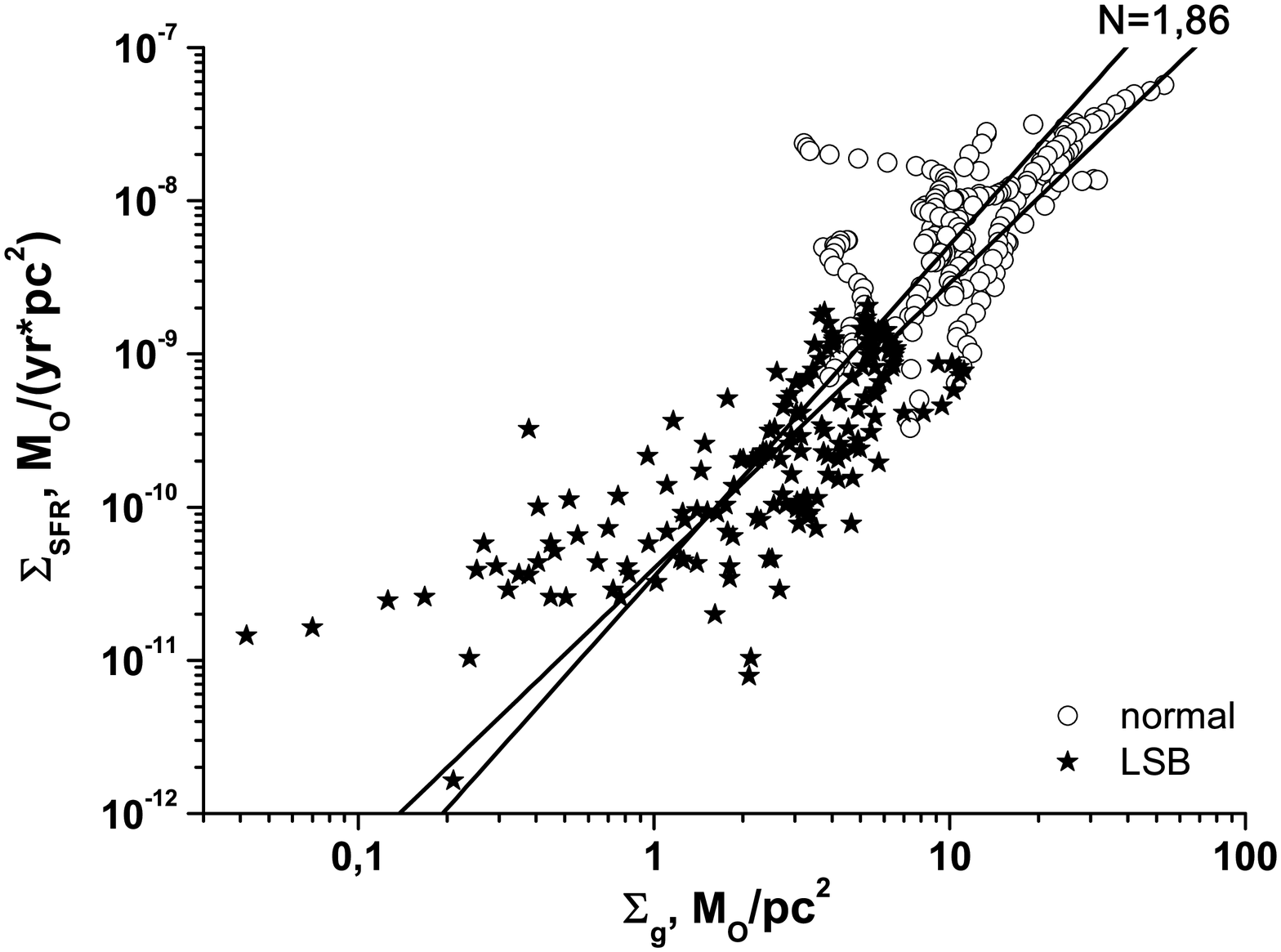}}\quad
\subfigure[]{\includegraphics[scale=0.30]{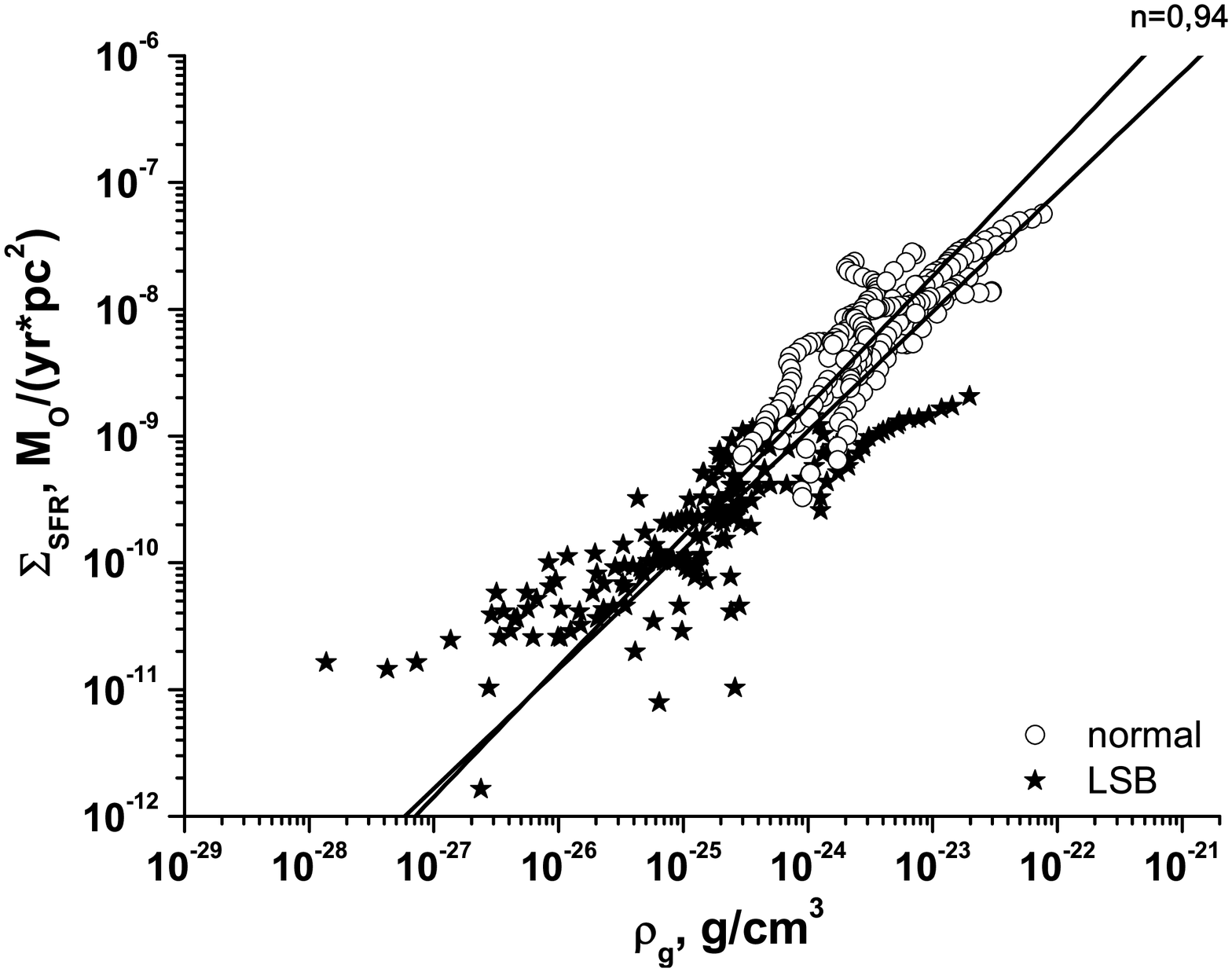}}} \caption{\footnotesize
$\Sigma_{SFR}$ plotted against surface gas density $\Sigma_g$ (a) and volume midplane gas
density $\rho_g$ (b). N and n are the coefficients of bisectors of regression
lines.}\label{fig2}
\end{figure}

In Figures~\ref{fig2}a,b the $\Sigma_{SFR}$ plots as a function of the surface gas density
$\Sigma_g=1.4\,(\Sigma_\textrm{HI}+\Sigma_{\textrm{H}_2})$ (a) and the volume gas density
$\rho_g$ (b), calculated as described above for normal and LSB galaxies.
%%++ Figure:2

A comparison of diagrams (a) and (b) clearly illustrates that $\Sigma_{SFR}$ correlates
with the volume gas density better than with the surface one: correlation coefficients $R$
= 0.85 and 0.91 correspondingly, or 0.85 and 0.94 if to exclude LSB galaxy F568-3 which
has the outstandingly low $\Sigma_{SFR}$ for its gas density (see Fig~\ref{fig2}(b)). The
most essential is that the dependencies for normal and LSB galaxies overlap while
continuing each other. It means that the same density of gas provides similar star
formation rate both in the outer regions of normal spiral galaxies and in the inner, most
gas-rich regions of LSB galaxies. A bisector between the regression lines,  found by the
least square method, has the slope $n$ which is close to unit (see Fig.~\ref{fig2}(b)).
Note that the most rarified regions of LSB galaxies
($\rho_g<10^{-26}~\textrm{g}/\textrm{cm}^3)$ spread very loose on the diagram. In the
mean, their $\Sigma_{SFR}$ is higher than expected. Disc overstability cannot explain it
because it would shift the points in the opposite direction. It seems, that the axial
symmetry and dynamical equilibrium may not be valid for the strongly rarified outer discs.

\begin{figure}
\centering
\mbox{\subfigure[]{\includegraphics[scale=0.40]{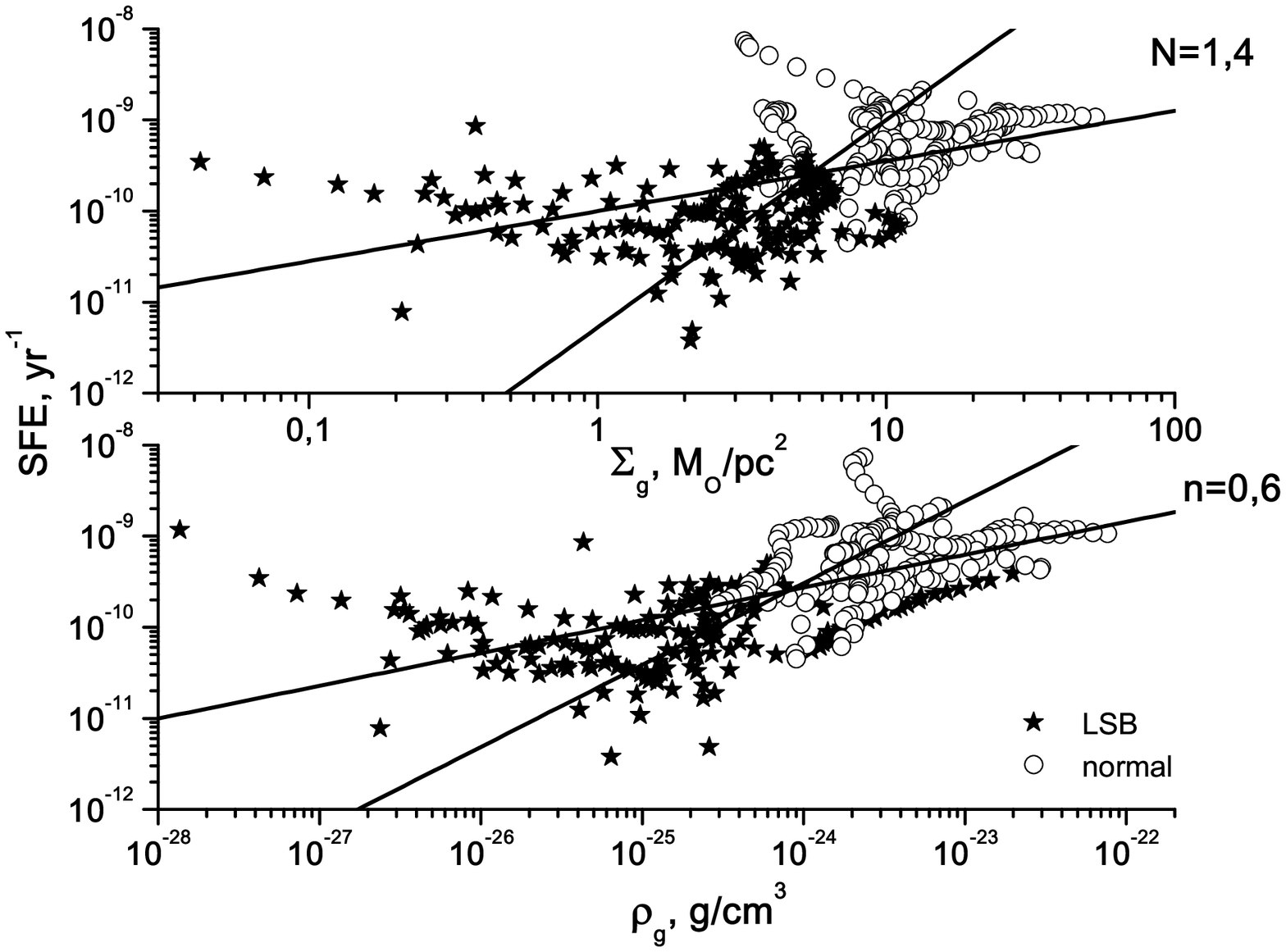}}\quad
\subfigure[]{\includegraphics[scale=0.40]{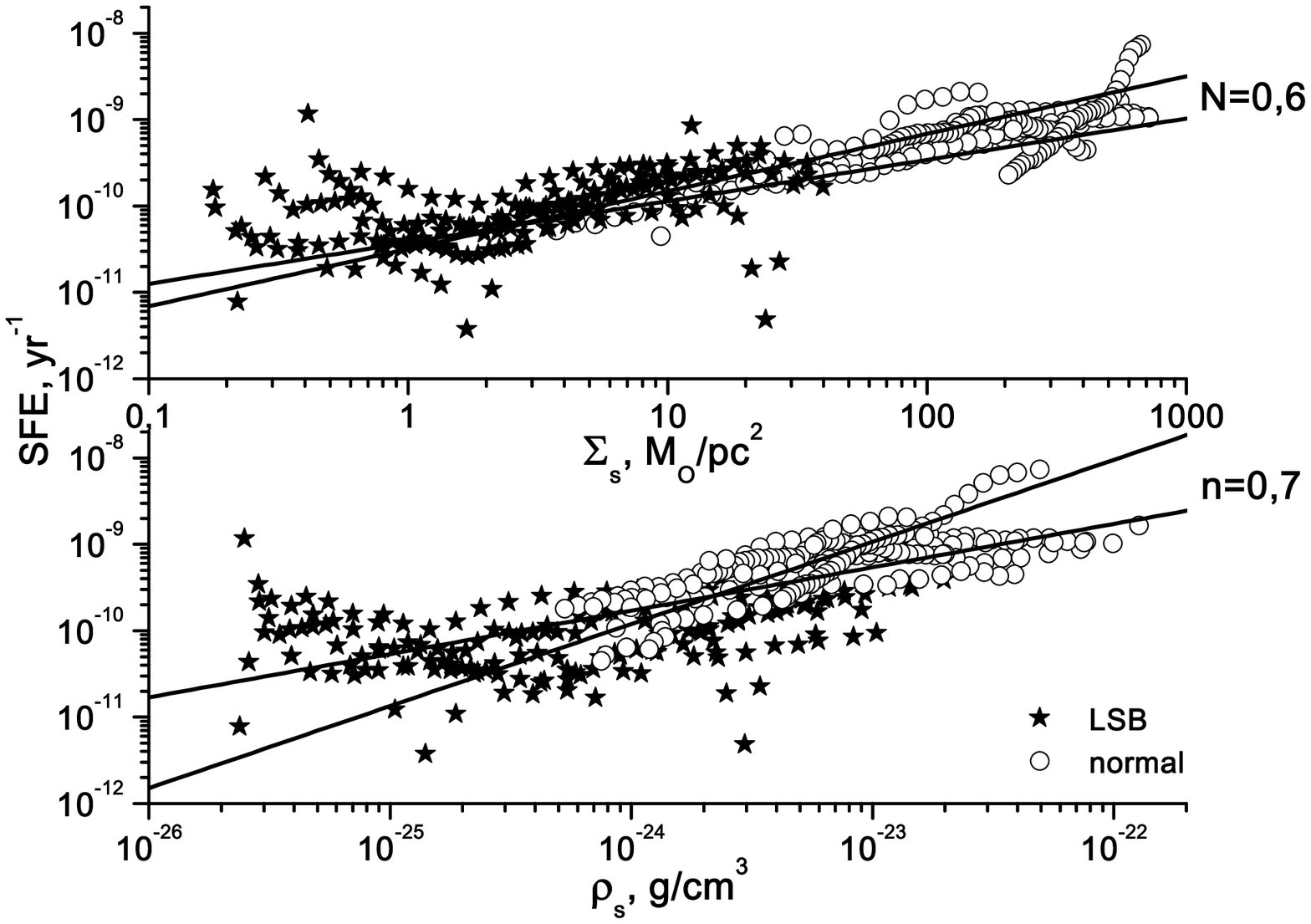}}}
\caption{\footnotesize $SFE$ plotted against gas densities $\Sigma_g$ and $\rho_g$ (a) and
stellar disc densities  $\Sigma_s$ and $\rho_s$ (b). N and n are the coefficients of bisectors
of regression lines.}\label{fig3}
\end{figure}

Now, to exclude the basic relationship between $\Sigma_{SFR}$ and gas content, we consider
the efficiency of star formation $SFE=\Sigma_{SFR}/\Sigma_g$. As observations show, $SFE$
is usually lower at the disc periphery than in its inner regions both for normal and LSB
galaxies (see f.e.~\cite{LeroyAll08, WyderAll09}). A key factor responsible for the
radial declining of $SFR$ may be a decreasing of gas or/and star volume density and gas
pressure, along the radius. However a correlation of $SFR$ with gas density looks very
loose (see Fig~\ref{fig3}(a).
%%++ Figure:3
The midplane gas pressure $P_g\sim\rho_g V_g^2$ should follow $\rho_g$(R) due to slow
radial variation of gas velocity dispersion $V_g$, hence both $\rho_g$ and $P_g$ may hardly
be the main factors which determine $SFE$.

Surprisingly, $SFE$ correlates more tightly with the stellar densities of discs -- both the
surface and volume ones (Fig~\ref{fig3}(b)). A connection between the disc stellar surface
densities and star formation rate for HSB spiral galaxies was first noted by Ryder and
Dopita~\cite{RyderDop94} and later confirmed by other authors
(\cite{ZasAbr06,LeroyAll08}). It appears that the there exist a single relationship
between $SFE$ and  $\Sigma_s$ or  $\rho_s$ for both HSB and LSB galaxies covering the
range $\Sigma_s=3-300 M_\odot$/$pc^2$. A comparison of diagrams in the top and in the
bottom of Fig~\ref{fig3}(b) shows that the correlation may be tighter for the surface
stellar disc density than for the volume one, but it may resulted from the different
methods of density estimations: $\Sigma_s$ is directly obtained from observational data,
whereas $\rho_s$ is the result of model calculations. A relationship $SFE(\rho_s)$ closely
follows the simple law $SFE\sim\rho_s^{1/2}$ or, in general, $(\rho_g+\rho_s)^{1/2}$ (not
shown here). Note that a root square of total disc density is proportional to the inverse
dynamical time, or to a frequency of crossing the midplane by any disc particle (gas
cloud). This oscillation frequency may serve as the universal factor which regulate $SFR$
equally well in high and low density regions if other factors are favorable for star
formation.

Another more sophisticated explanation of the nearly root-square relationship follows from
the  model of self-regulated star formation proposed by Ostriker et
al.~\cite{OstrikerAll10}. In this model a role of key factor for star formation plays UV
radiation created by young stars which keeps the thermal pressure of diffuse fraction of
gas at the level imposed by vertical force balance. For the diffuse gas-dominated regions
the model predicts $\Sigma_{SFR}\sim\Sigma_g\rho_{tot}^{1/2}$, where $\rho_{tot}$ is the
total midplane density of matter.

The most rarified outer regions of LSB galaxies ($\Sigma_s\leq 3 M_\odot$/$pc^2$) hardly
reveal any connection of $SFE$ with the gas or stellar disc density at all. For these
 regions $SFE$ lays within the range $10^{-10}-10^{-11}yr^{-1}$, which is higher in the mean than it
is expected from the sequences which fits the higher gas density regions of HSB spiral
galaxies (Fig 3b). Star formation rate in these regions  is governed  by some other
physical factors but the stellar or gas densities. Extremely low level of star formation
means that it should be concentrated in a small number of starforming sites with a random
local excess of gas density, so it is hard to expect them to follow general correlations.
 Both stellar and gaseous discs with such low densities may be irregular and far from equilibrium
 state, which makes the model we use for the density estimations to be unacceptable.
Accretion flows, minor mergers or interactions may be responsible for local gas
concentrations, as for a great diversity of the observed properties of extremely rarified
regions.

\section{Conclusions}

\hspace{0.6cm} A comparison of star formation rate and star formation efficiency  with the
disc an gas midplane densities we found for a sample of HSB and LBG galaxies shows that:
\begin{itemize}
\item both types of galaxies follow a single sequence $\Sigma_{SFR}-\rho_{g}$, revealing the
common Schmidt law within a wide range of gas densities down to $10^{-25}-10^{-26}
g/cm^3$;
\item Star formation efficiency  depends on the stellar density $\Sigma_s$ or $\rho_s$ in a
similar way for both HSB and LSB galaxies. Hence for a given stellar and gas densities the
conditions of star formation are similar in the outer discs of spiral galaxies and in the
inner, most gas-rich regions of LSB galaxies;
\item The relationship between $SFE$ and $\rho_s$ is close to the law $SFE$ $\sim
\rho_s^{1/2}$ expected in the theoretical model of self regulated star formation developed
 by Ostriker et al \cite{OstrikerAll10}. The alternative explanation is to propose that
$SFE$ is proportional to a frequency of vertical oscillation of gas clouds around the disc
midplane;
\item $SFR$ and $SFE$ in the most rarified regions of LSB galaxies ($\rho_g \leq 10^{-26}
g/cm^3$) do not reveal significant correlations with gas and stellar disc densities; the
model of equilibrium axisymmetrical disc is hardly applicable for them. Local density
fluctuations rather than azimuthally averaged disc parameters should play a crucial role
in the triggering of star formation there.
\end{itemize}

\end{document}